\documentclass[fleqn,10pt]{wlscirep}

\usepackage[utf8]{inputenc}
\usepackage[T1]{fontenc}

\usepackage{color}
\usepackage{amssymb}
\usepackage{amsmath}
\usepackage{bm}
\usepackage{graphicx}
\usepackage{kotex}
\usepackage{diagbox}
\usepackage{makecell}
\usepackage{multirow}
\usepackage{siunitx}
\usepackage{grffile}
\usepackage{float}
\usepackage{tabularx}

\makeatletter

\newcommand{\Rmnum}[1]{\expandafter\@slowromancap\romannumeral #1@}

\makeatother

\title{Engel's law in the commodity composition of exports}
\author[1]{Sung-Gook Choi}
\author[1,*]{Deok-Sun Lee}
\affil[1]{Department of Physics, Inha University, Incheon 22212, Korea}
\affil[*]{deoksun.lee@inha.ac.kr}
\date{\today}

\begin{abstract}
Different shares of distinct commodity sectors  in production, trade, and consumption illustrate how resources and capital are allocated and invested.   Economic progress has been claimed to change the share distribution in a universal manner as exemplified by the Engel's law for the household expenditure and  the shift  from primary to manufacturing and service sector in the three sector model. Searching for large-scale quantitative evidence  of such  correlation, we  analyze the gross-domestic product (GDP)  and international trade data based on the standard international trade classification (SITC) in the period 1962 to 2000. Three categories, among ten in the SITC, are found to have their export shares  significantly correlated with the GDP over countries and time; The machinery category has positive and  food and crude materials have negative correlations. The export shares of commodity categories of a country are related to its GDP by a power-law with the exponents  characterizing the GDP-elasticity of their export shares.  The distance between two countries in terms of their  export portfolios is measured to identify several clusters of countries sharing similar portfolios in 1962 and 2000. We show that  the countries whose GDP is increased significantly in the period are likely to transit to the clusters displaying large share of the machinery category. 
\end{abstract}

\begin{document}
\flushbottom
\maketitle 
\thispagestyle{empty}

\section*{Introduction}

From the consumers behaviors to  the sectoral composition of production, the economic structure of a system appears to be correlated with the  development level~\cite{kuznets1971economic,Pasinetti1983-PASSCA}.  Engel's law~\cite{engel1857} states that the expenditure on food falls as the household income increases~\cite{houthakker1957,RePEc:cpr:ceprdp:12387}. On a larger scale, working populations move  from primary (agricultural) through secondary (manufacturing) to tertiary (service) sector as economic progress is made~\cite{clark1940}. The share of the manufacturing sector in product increases with increasing gross domestic product (GDP) per capita across country and time, which has been recognized as a major component of the development pattern~\cite{kuznets1971economic,10.2307/1926806}.  These findings commonly suggest the presence of a economic growth {\it law}, similarly to that in the cellular network of living organisms; the mass abundances of distinct protein sectors change with the growth rate of a cell in different manners depending on the growth-limiting factors of the environment~\cite{doi:10.1146/annurev.mi.03.100149.002103,Scott1099,Hui784}.

Here we study the structure of international trade of individual countries searching for its development pattern  in the trade data over about 40 years. International trade is important for the structural transformation of a country's and the global economy~\cite{Chenery86book}, and  affected by the relationship between countries in political, social, cultural, and geographical contexts as well as their economic power~\cite{10.2307/1884452}.  The global organization of the international trade relationship between countries has been extensively investigated, revealing its far-from-random topological features and correlation with the GDP of  countries~\cite{PhysRevE.79.036115,PhysRevE.68.015101,PhysRevLett.93.188701,cha2010,1367-2630-17-1-013009,PhysRevE.95.052319}.  The trade data  also disclose the sectoral composition of the export of individual countries, which may hint at how economic growth and development are made at the level of individual countries by tuning resource allocation and export over different sectors~\cite{Hidalgo482,Ricardo-Hausmann:2013aa,khoslaexportmap,1742-5468-2017-12-123403} and a growth law in international trade.

How much values of individual commodities are exported by a country, which we  call its export commodity portfolio, is constrained by geography and production factor endowment~\cite{doi:10.1177/016001799761012334,doi:10.1177/016001799761012280,doi:10.1093/jeg/lbv028}. In this globalization era,  however, such constraints are increasingly weakened and goods tend to be manufactured by combining labor, capital, and technology across countries~\cite{Hummels200175}. Thus a country is allowed to control freely its resource allocation and investment to achieve maximum economic growth, which increases the  possibility of identifying the development patterns  in the commodity  composition of trade~\cite{linder1961essay,RePEc:eee:ecmode:v:23:y:2006:i:6:p:978-992,10.1257/aer.89.3.379,Tacchella:2018aa}.  Hajzler has investigated the trade data in the period 1970-1992 to find the decrease of the share of primary-sector commodities with time and also with the GDP per capita~\cite{hajzler03}, a model of which was also proposed in ~\cite{ECHEVARRIA2008192}.  

Previous studies of the spatial and temporal variation of the structure  of production and trade are, however,  based on a highly coarse-grained classification, the three-sector model~\cite{clark1940}; primary, manufacturing, and services. Much refined classification schemes are currently available, such as the Harmonized Commodity Description and Coding System (HS)~\cite{hs} or the standard international trade classification(SITC)~\cite{sitc4}, which should show the development pattern at a refined level of commodity classification. Here we analyze the trade data~\cite{NBERw11040} based on the SITC to investigate the shares of the refined commodity categories in the exports of about hundred countries in the period 1962 to 2000.  Depending on the number of digits we keep in the SITC code,  the number of commodity categories varies  from ten to hundreds.  

We find that among 10 categories at the first-digit level,  only one category, {\it Machinery and transport equipment}, belonging to the manufacturing sector, has its share in export  increasing significantly and persistently with the GDP of countries characterized by a positive scaling exponent every year. Two categories, {\it Food and live animals} and {\it Crude materials}, belonging to the primary sector, display negative correlations significantly  with the GDP.  The remaining categories  show little correlation.  The temporal variation  of the export share of the three categories also show correlations with the temporal variation  of the GDP.  The evolution of the whole export portfolio of countries also reveal similar correlation with the variation of the GDP. Given the wide variation of the portfolio of individual countries due to the influences of different environment and production factors, we extract several clusters of countries sharing similar export portfolios in 1962 and 2000 and show that  the countries whose GDP increase with time are likely to have transited  to a cluster displaying large share of the {\it Machinery} category in the export portfolio.  These results can be useful for understanding and designing the commodity composition of export of individual countries in the global economy. 

\section*{Results}
\subsection*{Preliminary: Commodity composition of world trade}

\begin{table*}
	\begin{tabular}{c|c|c}
		\hline\hline
		$p$ & Commodity category  &   GDP-elasticity of export share $\alpha_p$\\ 
		\hline
		0 & {\it Food and live animals} &  $-0.12\pm 0.06$  \\
		 1 & {\it Beverages and tobacco} &  $0.36\pm 0.18$ \\
		2 &{\it Crude materials} & $-0.09\pm 0.04$ \\
		3 &{\it Mineral fuels} & $-0.01\pm 0.23$  \\
		4 &{\it Animal and vegetable oils, fats and waxes} & $0.11\pm 0.09$  \\
		5&{\it Chemicals}  & $0.41\pm 0.14$  \\
		6 &{\it Manufactured goods classified by materials}& $0.31\pm 0.07$ \\
		7&{\it Machinery and transport equipment}& $0.67\pm 0.10$ \\
		8&{\it Miscellaneous manufactured articles}  & $0.38\pm 0.15$ \\
		9&{\it etc} & $0.25\pm 0.08$ \\
		\hline\hline
	\end{tabular}
	\caption{Ten commodity categories based on the first digit of the SITC, indexed by $p$, and their GDP-elasticity $\alpha_p$.}
	\label{table:SITC}
\end{table*}

From the NBER-UN data-set~\cite{NBERw11040}  and the GDP data compiled by Gleditsch~\cite{GLEDITSCH01102002} based on the Penn World Table 6.1~\cite{pwt61},  we obtain the export value  $F_p(c,t)$  of a commodity category $p$  for a country $c$,  and the GDP $W(c,t)$ in US dollars  for year $t$ with  $1962\leq t\leq 2000$.  See Methods for more details of compiling data-sets.  $112$ to $157$ countries are considered from year to year; Only considered are the countries having both the GDP and the export value available in the data-sets. The commodity classification is based on the  SITC4 Rev. 2 in which a 4 or 5-digit number is assigned to each category at the finest level~\cite{sitc4}. We use the first one or two digits to obtain a total of $10$ or $93$ categories, respectively, each denoted by $p$.  In Table~\ref{table:SITC}, $10$ commodity categories based on the first digit of the SITC are shown.  The SITC does not include services.  The categories with $p=0,1,2,3,4$ cover commodities related to raw materials or their elementary processing, belonging to the primary sector, and those with $p=5,6,7,8,9$ contain manufactured goods and more complex commodities than raw materials belonging to the manufacturing sector.

\begin{figure}
\includegraphics[width=10cm]{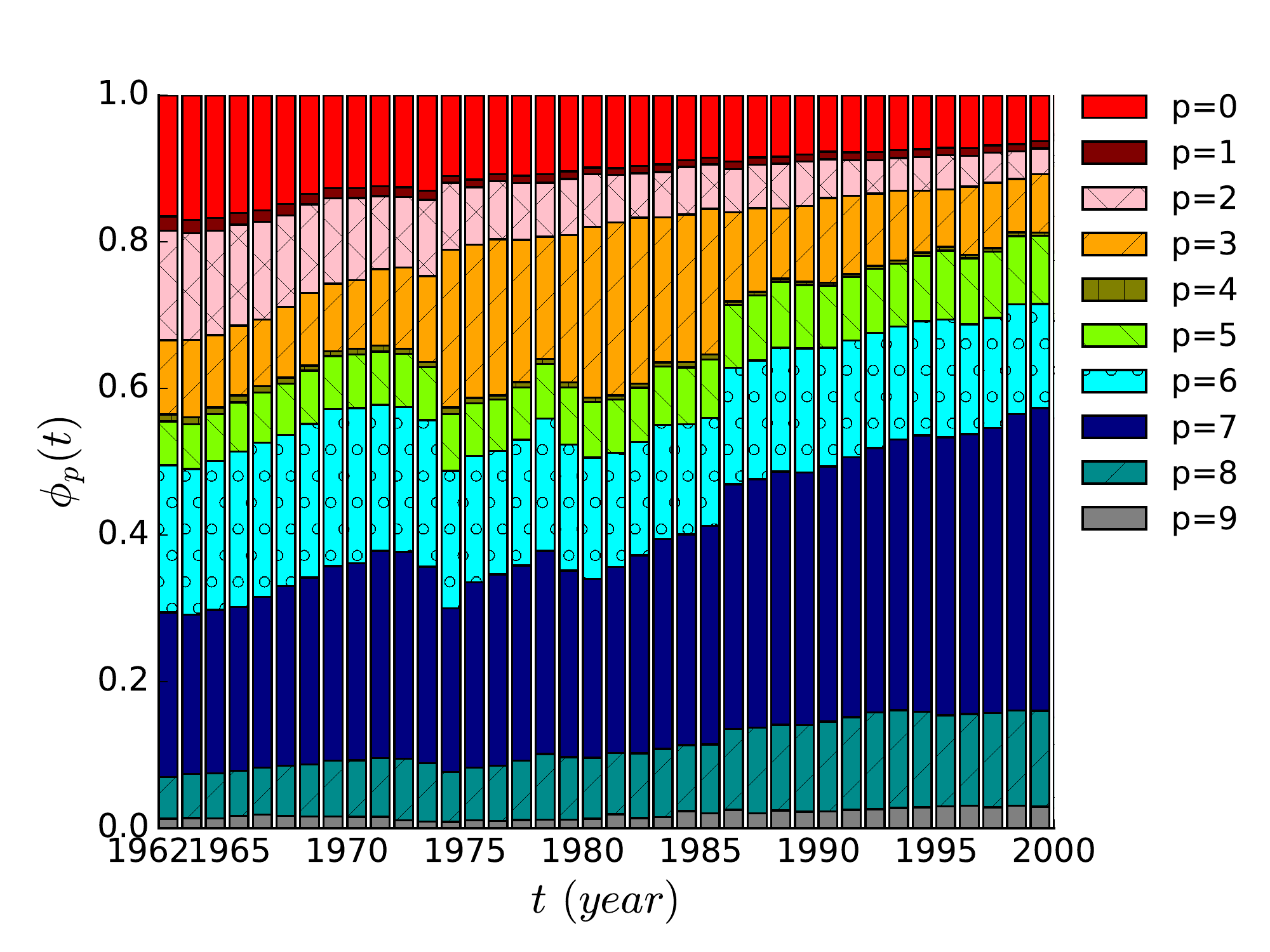}
\caption{The time-evolution of the global export share $\phi_p(t)$ of $10$ commodity categories in world trade.}
\label{fig:globalshare}
\end{figure}

The share of a category $p$ in the world trade is obtained by summing its export values over all countries and normalizing the sum as 
\begin{equation}
\phi_p(t) \equiv {\sum_c F_p(c,t)\over \sum_{c,p'} F_{p'}(c,t)},
\label{eq:phipt}
\end{equation}
which satisfies the relation $\sum_p \phi_p(t)=1$ for given year $t$. The global export shares of the ten commodity categories vary with time differently as shown in Fig.~\ref{fig:globalshare}. The category of {\it Food and live animals} ($p=0$)  and that of {\it Crude materials} ($p=2$) occupy $32\%$ in 1962, decreasing gradually with time  to $\phi_0(t) + \phi_2(t)=9\%$  for $t=2000$. On the contrary, the category of {\it Machinery and transport equipment} ($p=7$) and of {\it Miscellaneous manufactured articles} ($p=8$) find their shares doubled, increasing from $\phi_7(t) = 22\%$ and $\phi_8(t)=6\%$ for $t=1962$ to $\phi_7(t)=41\%$ and $\phi_8(t)=12\%$ for $t=2000$. Other categories do not show strong variation. The share of the primary sector $\phi_{\rm primary}(t) = \sum_{p=0}^4 \phi_p(t)$ decreases with time and that of the manufacturing sector $\phi_{\rm manufacturing}(t) = \sum_{p=5}^9 \phi_p(t)$ increases with time. 

Given that the gross world product $g(t)= \sum_c g(c,t)$ increases exponentially with time, the rise of the share of the manufacturing sector and the fall of the primary one in the considered period suggest the growth law of world trade: The global export share of manufacturing (primary) sector  increases (decreases) with the gross world product~\cite{hajzler03,ECHEVARRIA2008192}, recalling the law established for domestic production~\cite{kuznets1971economic,10.2307/1926806}, and Engel's law for the household expenditure. Then it is natural to ask whether there is a similar law governing the export shares of commodities of individual countries.   To answer this, we first present the commodity categories correlated with the GDP  in Sec.~\ref{sec:GDPcommodities}. The whole export portfolio of countries and its relation to the economic growth are studied in Sec.~\ref{sec:clustering}. 

\subsection*{GDP-correlated commodities}
\label{sec:GDPcommodities}

In this section we study how the share of each commodity category in the export of a country, called here {\it local} export share, varies with the GDP of the country. The equal-time cross-country correlation of the two quantities is investigated for each year and then temporal correlation is explored. 

To exclude the influence of the exponential growth of the export value and the GDP with time and concentrate on their endogenous variations, we consider the share of each commodity category in the export  and the normalized GDP of a country defined as 
\begin{equation}
\phi_p(c,t) \equiv {F_p(c,t) \over \sum_{p'} F_{p'}(c,t)},
\label{eq:phipct}
\end{equation}
and 
\begin{equation}
g(c,t) \equiv {W(c,t) \over \sum_{c'} W(c',t)},
\label{eq:gct}
\end{equation}
respectively, where $c$ is the index of a country. Note that $\sum_p \phi_p (c,t)=1$ for given $c$ and $t$ and  $\sum_c g(c,t)=1$ for given $t$.

\begin{figure}
\includegraphics[width=10cm]{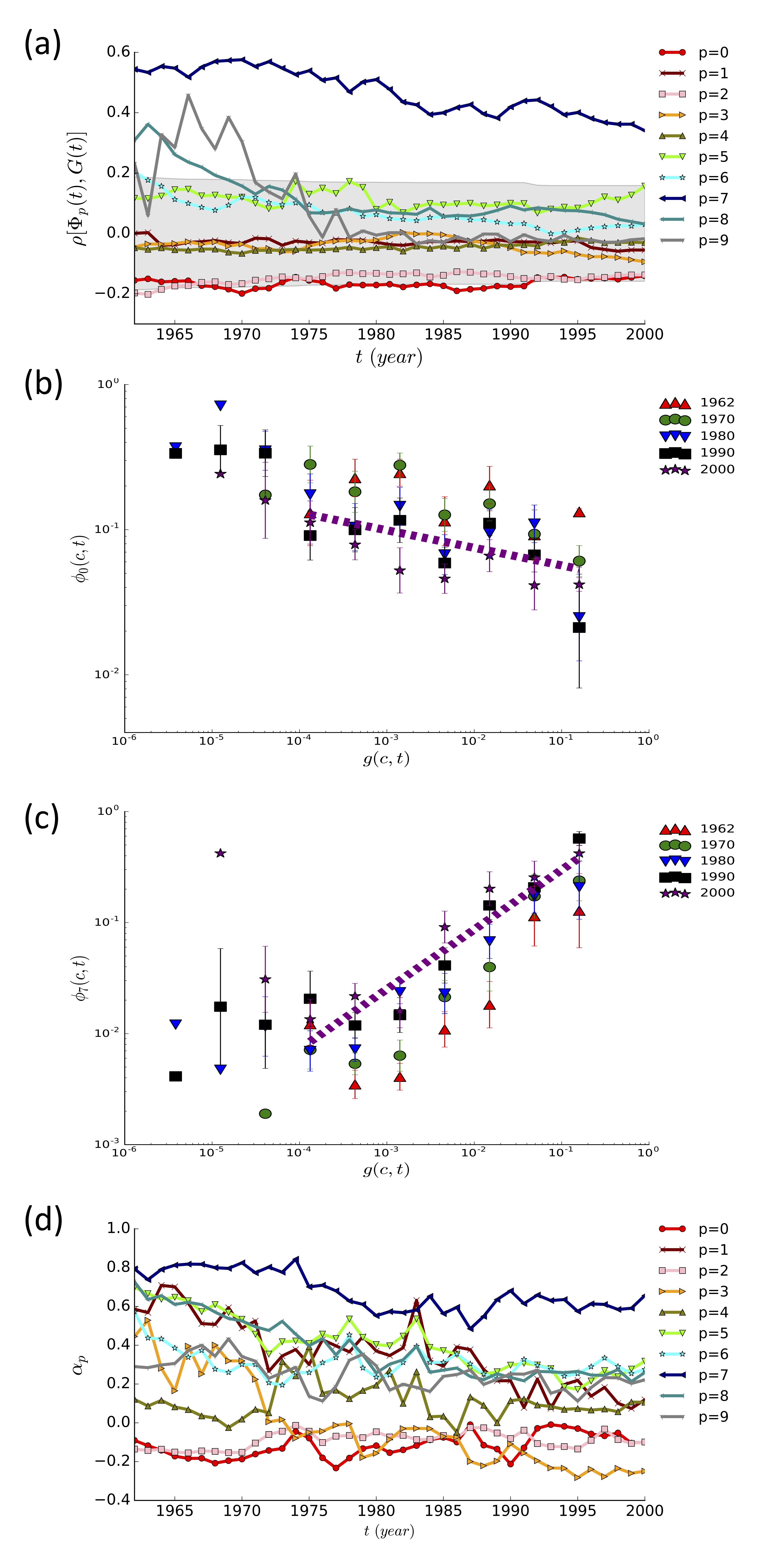}
\caption{
Correlation between local export share  and GDP.  Ten commodity categories based on the first digit of the SITC code are considered.
(a) The Pearson correlation coefficient in Eq.~(\ref{eq:rho}) between the local export share profile of a commodity and the  GDP  profile as a function of time $t$. The range of $\rho$ with P $>0.05$ is shaded. 
(b) Plots of the local export share $\phi_0(c,t)$ of the {\it Food and live animals}  category versus the normalized GDP $g(c,t)$ of countries  for selected years. The dashed line has slope $-0.12$. 
 (c) The same plot as (a) for the  {\it Machinery and transport equipment} category ($p=7$). The dashed line has slope $0.67$. 
 (d) GDP-elasticity $\alpha_p(t)$ in Eq.~(\ref{eq:alphadefined}).  
 }
\label{fig:corr_phi_g}
\end{figure}

\subsubsection*{Equal-time cross-country correlation between local export share and GDP}

To see how the local export shares of individual commodities relate to the GDP,  we  fix time $t$ and measure  the correlation between the export share profile  $\Phi_p(t) \equiv \{\phi_p(c,t)|c=1,2,\ldots\}$ of a category $p$ and the normalized GDP profile $G(t) \equiv \{g(c,t)|c=1,2,\ldots\}$ 
\begin{equation}
\rho[\Phi_p(t), G(t)]  =  {
\langle \phi_p(c,t) g(c,t)\rangle - \langle \phi_p(c,t)\rangle \langle g(c,t)\rangle
\over 
\sigma_{\Phi_p(t)} \sigma_{G(t)}},
\label{eq:rho}
\end{equation}
where $\langle \cdots \rangle$ is the average over countries having both  $g(c,t)$ and $\phi_p(c,t)$ available in the studied data-sets and $\sigma$ is the standard deviation. The correlation $\rho$ turns out to be quite different among  10 commodity categories as shown   in Fig.~\ref{fig:corr_phi_g} (a). While fluctuating with time,  $\rho$ remains negative for $p=0,1,2,3$, and $4$ and  positive for $p=5,6,7,8$.  The correlation for $p=9$ is positive until around 1980 and becomes  negative afterwards. These results are in agreement with known results - the fall of primary and the rise of manufacturing sector with the progress of a country's economy.

The correlation $\rho$  is not significant for all categories, however; It is only for $p=7$ that P value is less than $5\%$ for all $t$. For $p=0$ and $p=2$, P values are around $5\%$ for the considered period.  The export share of  the {\it Food and live animals} category and  of the {\it Machinery and transport equipment} category are shown as functions of the normalized GDP $g(c,t)$ for selected years in Fig.~\ref{fig:corr_phi_g} (b) and (c), respectively. We find that $\phi_p(c,t)$ for $p=0, 2$ or $p=7$ decays or grows algebraically with $g(c,t)$ over a wide range of $g(c,t)$ for every year $t$ as 
\begin{align}
\phi_p (c,t) \sim g(c,t)^{\alpha_p(t)}
\label{eq:alphadefined}
\end{align}
with the scaling exponent $\alpha_{0}\simeq -0.12\pm 0.06$, $\alpha_2\simeq -0.09\pm 0.04$, and $\alpha_{7}\simeq 0.67\pm 0.10$. The error arises from the fluctuation over time [Fig.~\ref{fig:corr_phi_g} (d)]. Such scaling behaviors are related to the {\it elasticity} in the economics context~\cite{varian_micro}, in which the elasticity of a quantity $A$ with respect to $B$ indicates the ratio of their relative variations ${A^{-1} \Delta A  \over B^{-1} \Delta B}$. Therefore the elasticity of the export share of a category $p$ with respect to the GDP is approximated by the scaling exponent $\alpha_p$ in Eq.~(\ref{eq:alphadefined}) 
\begin{equation}
 {\phi_p^{-1} \Delta \phi_p \over g^{-1} \Delta \, g} \simeq {d \ln \phi_p(c,t) \over d \ln g(c,t)}.
\end{equation}
We will call $\alpha_p$ in Eq.~(\ref{eq:alphadefined})  the GDP-elasticity of the local export share of commodity category $p$. A category $p$  with positive (negative) $\alpha_p$  has its share in the export of a country increasing (decreasing) with the economic power - GDP  of the country. Therefore the categories having negative $\alpha_p$ are similar to ``food'' the expenditure of which falls as the household income increases as stated by the Engel's law.  The {\it Machinery} category $(p=7)$ requires more complex and high-end technology, which are likely to be available and demanded in developed and rich countries, presumably underlying the positive elasticity seen  in Fig.~\ref{fig:corr_phi_g} (c).  The GDP-elasticities of all categories are shown in Table~\ref{table:SITC}.

When we use 93 commodity categories based on the first two digits of the SITC codes, we find consistently significant correlations between  export share and  GDP for 6 categories. They are 
{\it Artificial resins and plastic materials, and cellulose esters etc} ($p=58$), 
{\it Power-generating machinery and equipment} ($p=71$), 
{\it General industrial machinery and equipment} ($p=74$),
{\it Office machines and automatic data processing equipment} ($p=75$),
{\it Road vehicles} ($p=78$), 
and {\it Professional, scientific, controlling instruments, apparatus} ($p=87$). 
See Supplementary Information (SI) and Fig. S1.
All are positive correlations. Four belong to the {\it Machinery and transport equipment} ($p=7$) one-digit category.  We find no significant {\it negative} correlations  at the two-digit level, which suggests that  the driving force of economic development is acted to  increase the export share of selected commodity categories and the decrease of the share of the remaining categories is its passive outcome subject to randomness resulting in weak negative correlations with the GDP.

\begin{figure}
\includegraphics[width=10cm]{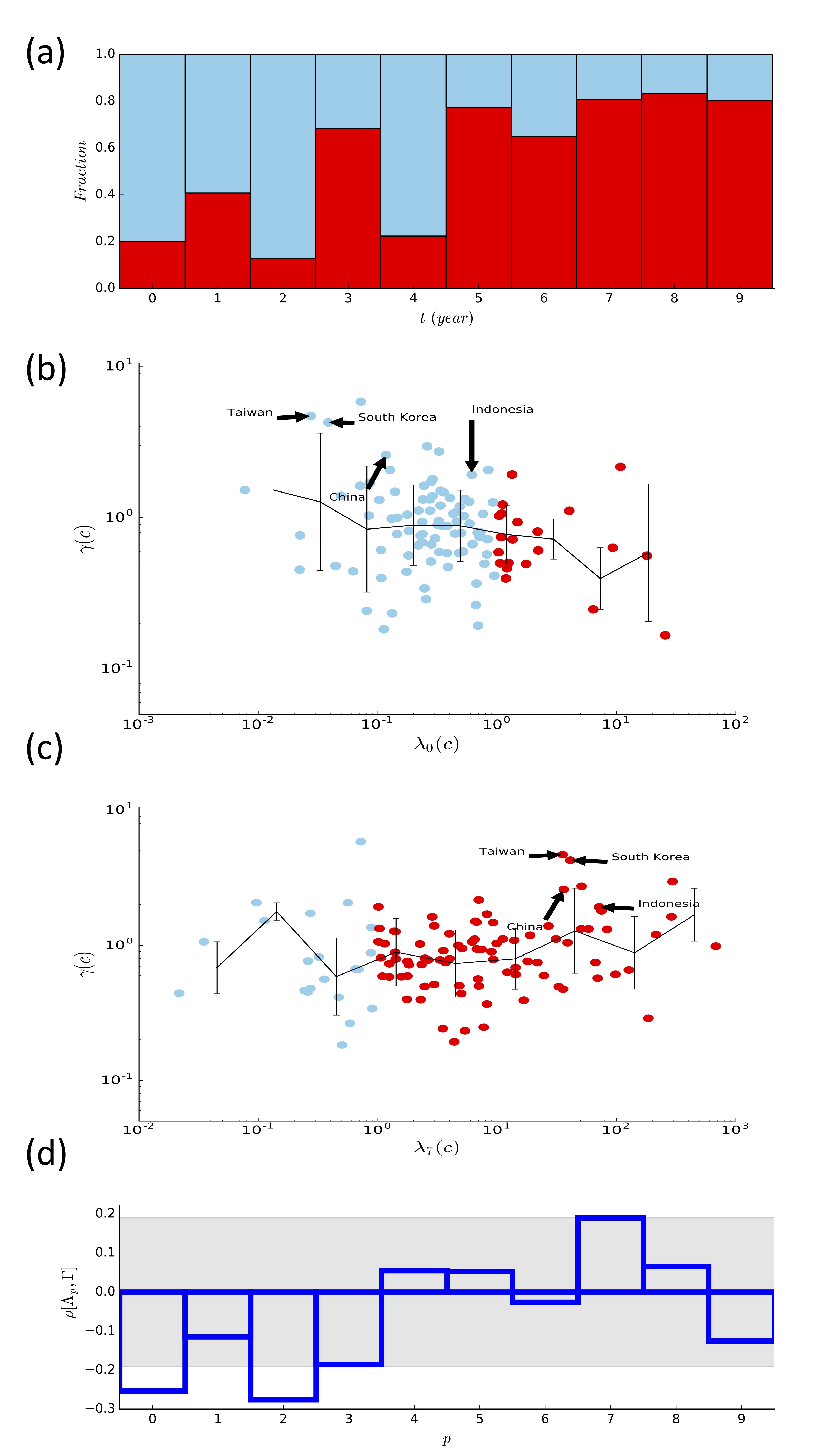}
\caption{
Relation between the multiplicative time variation of the local export share of commodities and that of the GDP from 1962 to 2000.  
(a) The fraction of countries having the share of category $p$ increased,  $\lambda_p(c)>1$ (red) and  decreased $\lambda_p(c)<1$ (blue).
(b) The multiplicative time variation $\lambda_0(c)$ of the export share of  the {\it Food and live animals} category $(p=0)$ as a function of the multiplicative time variation of   the GDP  $\gamma(c)$ of countries. Red or blue color represents whether $\lambda_p(c)$ is larger or smaller than one. The lines represent the geometric average and the errorbar is the standard deviation.  
(c) The same plot as (b) for the {\it Machinery} category ($p=7$). 
(d) The Pearson correlation coefficients between the multiplicative variation profile of the export share of each commodity category $p$ and of the GDP $\rho[\Lambda_p,\Gamma]$ with $\Lambda_p = \{ \lambda_p(c)|c = 1,2, \ldots\}$ and $\Gamma = \{\gamma(c)|c=1,2,\ldots\}$ 
are shown. The range of $\rho$ with $P>0.05$ is shaded. }
\label{fig:ratio}
\end{figure}

\subsubsection*{Multiplicative time variation of local export share of commodities}

Given the global trend pushing towards a larger share of the manufacturing sector both in product and trade, as shown in Fig.~\ref{fig:globalshare}, and the significant correlations between the GDP and the export shares of selected categories, we are led to wonder  how individual countries modify their export commodity portfolio   $\Phi(c,t) \equiv \{\phi_{p_1}(c,t), \phi_{p_2}(c,t),\ldots\}$  between $t=1962$ and $2000$. Will a country achieve economic growth if it increases the export share of the {\it Machinery} category and decrease the share of {\it Food} and {\it Crude materials} with time? 61  countries had  $\phi_0$ and $\phi_2$ decreased and $\phi_7$ increased in the period 1962 to 2000. Over $50\%$ of the countries increased the share of category $p=3,5,6,7,8,9$ and decreased the share of $p=0,1,2,4$ in 2000 with respect to 1962 [Fig.~\ref{fig:ratio} (a)].

We find that a change of the export share of the GDP-correlated commodities  leads to the growth of the GDP only when the change follows the global trend; The countries have their GDP increased if $\phi_0$ decreases and $\phi_7$ increases [Fig.~\ref{fig:ratio} (b) and (c)]. We consider the multiplicative time variation in the period 1962 to 2000 of the export share and the GDP defined as 
\begin{equation}
\lambda_p(c) = {\phi_p(c,2000)\over \phi_p(c,1962)}, \ \ \gamma(c) = {g(c,2000)\over g(c,1962)}.
\label{eq:clusterquantities}
\end{equation}
Among 10 commodity categories, a significant correlation between the multiplicative variations of export share $\{\lambda_p(c)|c=1,2,\ldots\}$ and that of GDP $\{\gamma(c)|c=1,2,\ldots\}$ is identified only for $p=0,2$, and $7$ with P value $\simeq 0.008, 0.004$ and $0.05$, respectively, as shown in Fig.~\ref{fig:ratio} (d). This suggests that the economic growth of individual countries in the studied period  may be attributed to the increase of the share  of the {\it Machinery} category ($p=7$) and the decrease of  the share   of the {\it Food and live animals} ($p=0$) and {\it Crude materials} ($p=2$) in their exports. In Fig.~\ref{fig:ratio} (b) and (c) are marked some of the countries having $\phi_0$ decreased, $\phi_7$ increased, and the GDP increased as examples. 

Identifying the three categories $p=0,2,$ and $7$ whose export shares are relevant to the GDP and its variation is one of the main results of the present study. This can be considered as Engel's law for the structure of export of individual countries. We can say that countries having large share of the {\it Machinery} category in export are likely to be richer than those having small share. Similarly, countries decreasing  the share of the {\it Crude material} category in export are likely to achieve economic growth and find their GDP increased  in 2000 with respect to 1962. Therefore the increase or decrease of these three  categories $p=0,2,$ and $7$ is an important ingredient of the development patterns of export.

\subsection*{Classification and transition of export commodity portfolios}
\label{sec:clustering}

The distributions of export share over categories which we call the export commodity portfolios look disordered and random~\cite{Ricardo-Hausmann:2013aa}, defying seemingly a simple law or pattern. Such complexity  can be understood, for the commodity composition of export  is determined by not only the efforts towards economic development but also the constraints and influences imposed by geography, factor endowment, culture, politics,  and international relationships. Nevertheless, given the GDP-correlated categories in addition to the weakening of the constraints due to globalization, we expect that the whole export portfolios of countries and their time evolution  may be analyzed in terms of their interplay with economic growth via the GDP-correlated categories, which can offer a useful framework for the development pattern of international trade.

\subsubsection*{Classification of export commodity portfolios in 1962 and 2000}

We begin with classifying the export commodity portfolios of individual countries, which reduces the vast space of all possible  export portfolios.  Our idea is to group the countries having similar export portfolios into clusters and investigate the {\it average} portfolio of each obtained cluster,  the evolution of which is of interest to us particularly in connection with the variation of the GDP of the countries belonging to the cluster. 

\begin{figure*}
\includegraphics[width=17cm]{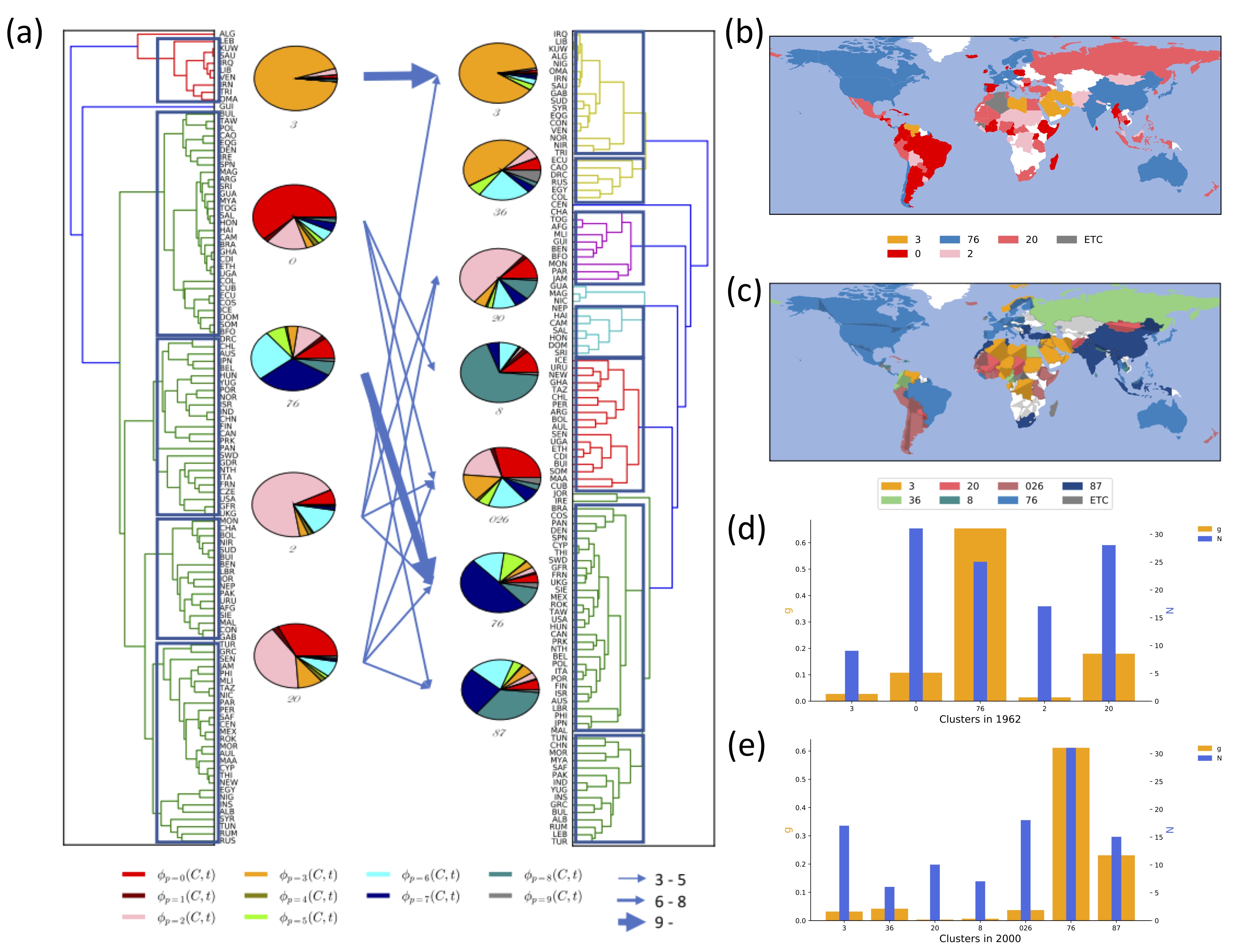}
\caption{
Clusters of countries sharing similar export portfolios.  
(a) Dendrogram of countries based on the Euclidean distance $d$ of their export portfolios in 1962 (left) and in  2000 (right). Clusters of countries are obtained by removing links of $d>d_c = 0.45$ in the dendrogram and their portfolios are represented in pie charts.  Each  arrow from a cluster in 1962 to one in 2000 represents the transition of countries between the clusters and the arrow width increases with the number of  countries.   No arrow is drawn for  the transition of just one or two countries. 
(b) Geographic distribution of countries belonging to each cluster in 1962. Countries are colored according to their clusters. 
(c)  The same as (b) for year 2000. 
(d) The number of countries and the  normalized GDP of  each cluster in 1962. 
(e) The same as (d) for year 2000.}
\label{fig:Cluster}
\end{figure*}

The dissimilarity of the export  portfolio between two countries is here quantified by the Euclidean distance  $d(c_1,c_2,t) \equiv \sqrt{\sum_p (\phi_p(c_1,t) - \phi_p (c_2,t))^2}$ with $p$  the two-digit SITC code. Linking iteratively countries and their clusters in the increasing order of the distance, we obtain two dendrograms given in Fig.~\ref{fig:Cluster} (a) for $t=1962$ and $t=2000$. The distance between clusters grows gradually with the iteration  until $d\simeq 0.45$  and fast afterwards for both years, and therefore we remove the links connecting clusters farther from one another than threshold $d_c = 0.45$. This leads  to   five clusters   in  1962 and seven clusters in 2000 [Fig.~\ref{fig:Cluster} (a)]. For each cluster of countries $C=\{c_1, c_2,\ldots\}$,   we obtain the export share  $\phi_p(C,t)$ of category $p$ and the normalized GDP evaluated as 
\begin{equation}
\phi_p(C,t) = {\sum_{c\in C} F_p(c,t) \over \sum_{c\in C} \sum_{p'} F_{p'} (c,t)}, \ \ g(C,t) = \sum_{c\in C} g(c,t).
\label{eq:phipCt}
\end{equation}
The export portfolio $\Phi(C,t) = \{\phi_{p_1}(C,t), \phi_{p_2} (C,t),\ldots,\}$ of each cluster $C$ is represented as a pie chart  in Fig.~\ref{fig:Cluster} (a). 

The portfolios of the obtained clusters are characterized commonly by one or a few dominant commodity categories. 
See SI and Table S1.
We name each cluster by the most dominant categories the summed share of which just exceeds 50\%; We have five clusters named $\mathit{0}$, $\mathit{20}$, $\mathit{2}$, $\mathit{3}$, and $\mathit{76}$  in 1962 and seven clusters named $\mathit{026}$, $\mathit{20}$, $\mathit{3}$, $\mathit{36}$, $\mathit{76}$, $\mathit{87}$, and $\mathit{8}$ in 2000. The portfolios of these clusters are quite different. For instance, the cluster $\mathit{026}$ of 18 countries in year 2000 has $\phi_0 =0.28, \phi_2 = 0.19, \phi_6 = 0.16$ while the $\mathit{3}$ cluster displays $\phi_3 = 0.93$ and  $\phi_3=0.88$ respectively  in 1962 and  2000.   In 1962, most of these dominant categories belong to the primary sector.

One can use the Euclidean distance of the portfolios between the clusters of 1962 and 2000 to pair clusters of 2000 and of 1962. We find  that the clusters $\mathit{3}$, $\mathit{20}$, $\mathit{76}$, $\mathit{20}$, and $\mathit{026}$ in 2000 are closest to the clusters $\mathit{3}$, $\mathit{20}$, $\mathit{76}$, $\mathit{2}$, and $\mathit{0}$ in 1962, respectively. The cluster $\mathit{20}$ of 2000 is found to be almost equally close to both $\mathit{20}$ and $\mathit{2}$ of 1962. 
See SI and Table S2.
for the distance of  every pair of clusters of 1962 and 2000. 

 The countries in some clusters tend to be clustered geographically  as  seen in Fig.~\ref{fig:Cluster} (b) and (c), demonstrating the importance of the geographic environment and factor endowment.  For instance, Oman, Saudi Arabia, Kuwait, Iraq, and Iran are close to one another located in the Middle East and belong to the cluster $\mathit{3}$ in both years displaying $93\%$ and $88\%$ of their export in the {\it Mineral fuels} category $(p=3)$.  In 1962,  many  countries of the cluster $\mathit{0}$ are in South America and the cluster $\mathit{2}$ and $\mathit{20}$ are in Africa. On the other hand, the cluster $\mathit{76}$ of 1962 are dispersed across North America, Western Europe, and East Asia. The countries in the clusters $\mathit{76}$ and $\mathit{87}$ in 2000 are distributed world wide. 
 
The clusters also show big difference in the economic power. We present the number of belonging countries  and the sum of their normalized GDPs, given in Eq.~(\ref{eq:clusterquantities}), of each cluster in Fig.~\ref{fig:Cluster} (d) and (e) for 1962 and 2000, respectively. The GDP   is  especially high for the cluster $\mathit{76}$ in 1962 and $\mathit{76}$ and $\mathit{87}$ in 2000 even if  their large numbers of countries are counted. In contrast, the GDP of  the cluster $\mathit{20}$ in 2000 is quite low.  Such different GDPs of the clusters can be attributed to their different shares of  the GDP-correlated categories.

\subsubsection*{Transition of the export commodity portfolio}

\begin{table*}
	\begin{tabular}{c||c|c|c|c|c|c|c}
	   \diagbox{$C^{(1962)}$}{$C^{(2000)}$}& $\mathit{026}$  & $\mathit{20}$ & $\mathit{3}$ & $\mathit{36}$ & $\mathit{76}$ & $\mathit{87}$ & $\mathit{8}$ \\
	     		\hline\hline	
		$\mathit{0}$ &0.53 (8) &0.71 (2) &0.46 (1) & 1.1 (3) & 1.1 (6)& 0.35 (2)& 0.97 (6)\\
		$\mathit{20}$&0.82 (6) &0.80 (3) & 0.70 (2)& 0.42 (2)& 1.8 (5)& 1.2 (8)& \\
                 $\mathit{2}$ & 0.55 (3)& 0.55 (4)&0.55 (4) & & 2.3 (3)& 2.0 (1)& 1.0 (1)\\
                 $\mathit{3}$ & & & 0.89 (8) & & & 0.49 (1)&\\
                 $\mathit{76}$ & 1.2 (1) & &0.88 (1)&0.17 (1) & 0.88 (17)& 1.9 (3)& \\		
	\end{tabular}
	\caption{
	The multiplicative time variation $\gamma(C^{(1962)}\to C^{(2000)}) = {g(C^{(1962)}\to C^{(2000)},2000) \over g(C^{(1962)}\to C^{(2000)},1962)}$ of the normalized GDP of the countries transiting from a cluster  of 1962 to the one of 2000. Rows and columns correspond to the cluster of 1962 and that of 2000. The number of countries for each transition is shown in parenthesis and empty space means no country for the corresponding transition.}
	\label{table:transition}
\end{table*}

\begin{figure}
\includegraphics[width=10cm]{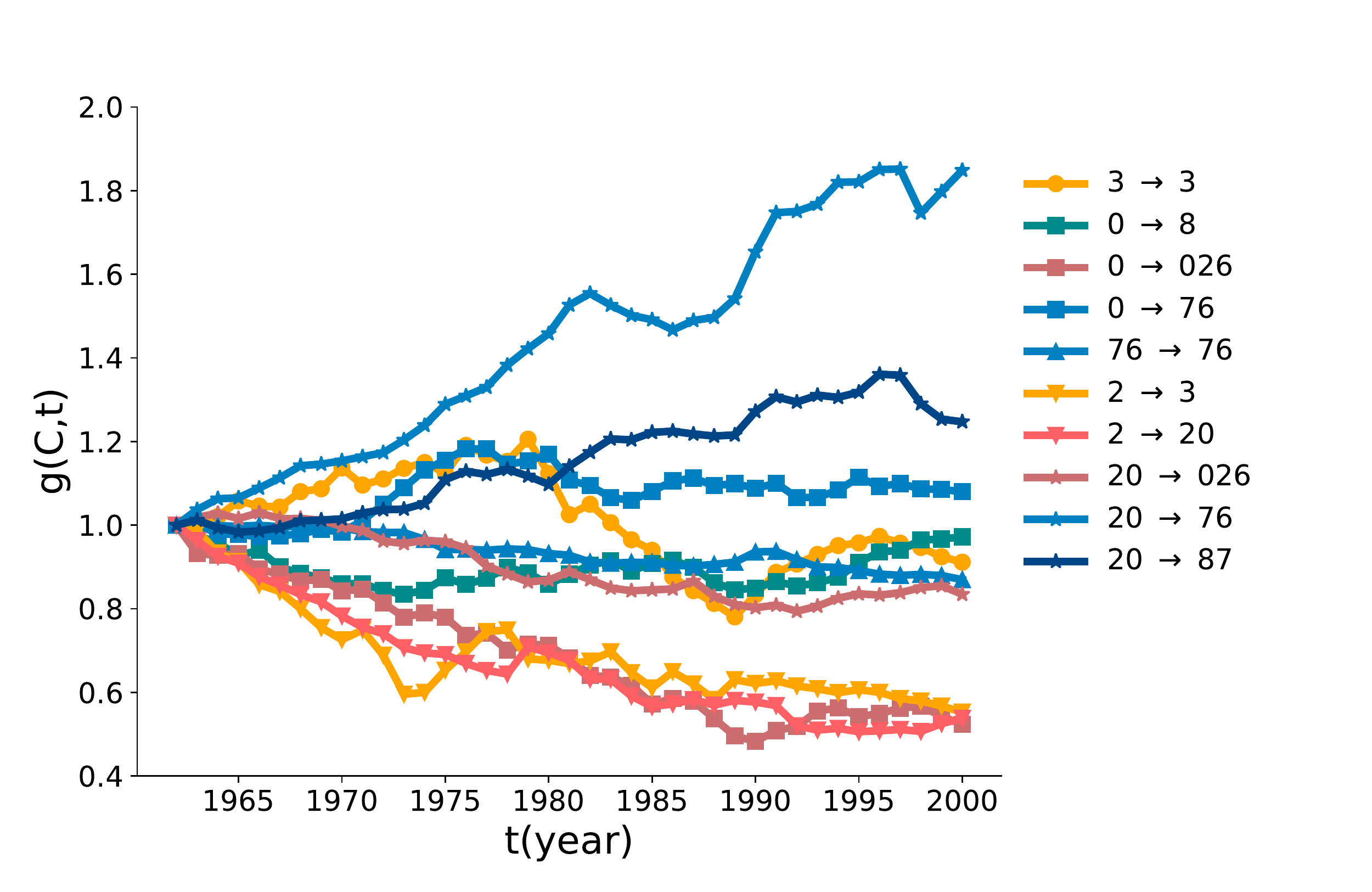}
\caption{
The normalized GDP $g(C^{(1962)} \to C^{(2000)}, t)$ of the countries transiting from a cluster $C^{(1962)}$ of 1962 to the one $C^{(2000)}$ of 2000 as a function of time $t$. The transitions of no more than 3 countries are excluded. The shape of data points varies with the cluster of 1962 and the color varies with the cluster of 2000.}
\label{fig:GDPallyear}
\end{figure}

With the obtained portfolio-based clusters of countries, one can take a coarse-grained view of how countries modify the export portfolios and  how it relates to economic growth. Some countries maintain their portfolios and remain in the qualitatively same cluster.  Eight out of nine countries in the cluster $\mathit{3}$ of 1962  are found again in the  cluster $\mathit{3}$ of 2000 except for  Lebanon. 17 out of 23 countries, mostly developed ones, stay in cluster $\mathit{76}$ while the share of category $p=7$ in the cluster is much increased in 2000 compared with 1962.  

Many countries  transit between different clusters implying  significant modifications of their export portfolios. The countries belonging to the clusters $\mathit{0}$, $\mathit{2}$ and $\mathit{20}$ of 1962 are found to be relatively evenly dispersed over multiple clusters in 2000 as shown in Fig.~\ref{fig:Cluster} (a) and Table~\ref{table:transition}. Most interestingly, economic growth is differentiated by the transition between clusters. For example, Malaysia (Pakistan) transits from $\mathit{2}$ to $\mathit{76}$ ($\mathit{87}$) and finds its GDP increased from $g=0.002 (0.003)$ to $g=0.005 (0.007)$, which is a fast growth with respect to the multiplicative time variation of all 17 countries, $\gamma(\mathit{2}^{(1962)}) = 0.96$, belonging to $\mathit{2}$ of 1962. Both South Korea and Thailand transit from $\mathit{20}$ to $\mathit{76}$ and their GDP  increases $4.27$ and $3.0$ times, respectively. Indonesia transits from $\mathit{20}$ to $\mathit{87}$ and its GDP is doubled. Taiwan moves from $\mathit{0}$ to $\mathit{76}$ with GDP increased 4.7 times. 

These examples reveal a common feature: Transition to a cluster $\mathit{76}$ or $\mathit{87}$ in 2000, consistent with the global trend,  is likely to be necessary for economic growth.  Note that those clusters are characterized by large shares of category $p=7$ and $6$ or $8$.   To check this systematically, we compute the multiplicative time variation of the normalized GDP of the countries transiting from a cluster $C^{(1962)}$ of 1962 to the one $C^{(2000)}$ of 2000
\begin{equation}
\gamma(C^{(1962)}\to C^{(2000)}) = {g(C^{(1962)}\to C^{(2000)},2000) \over g(C^{(1962)}\to C^{(2000)},1962)},
\label{eq:gammatransition}
\end{equation}
where $g(C\to C',t) = \sum_{c\in C\to C'} g(c,t)$ is the sum of the GDPs of the countries transiting from $C$  to $C'$, and present the result in Table~\ref{table:transition}.
Among the transitions taken by more than one countries, the transition from $\mathit{0}$ to $\mathit{36}$ or$\mathit{76}$, from $\mathit{20}$ to $\mathit{76}$ or $\mathit{87}$, from $\mathit{2}$ to $\mathit{76}$, and from $\mathit{76}$ to $\mathit{87}$ are associated with the increase of their GDP.  

The time-evolution of the  GDP over the whole period for each transition groups of more than three countries are presented in Fig.~\ref{fig:GDPallyear}, which shows the persistent increases of GDP for the countries transiting to $\mathit{76}$ and decrease for those to $\mathit{3}$, $\mathit{20}$ or $\mathit{026}$.  These results imply that the modification of the export portfolio plays an important role in economic growth. Yet there are exceptions like oil-producing countries staying in the cluster $\mathit{3}$, the normalized GDP of which neither decreases or increases significantly;  Their GDP highly depends on the world market situation of petroleum. The time-evolution of the GDP for all cases of transitions  is shown in Fig. S2. 

Among a total of 18 countries whose GDP has been increased 1.5 times or more in the period 1962 to 2000, seven countries are oil producers belonging to clusters $\mathit{3}$ or $\mathit{36}$ in 2000, and nine countries are in clusters $\mathit{76}$ or $\mathit{87}$ in 2000, including  the newly industrialized countries(NICs)~\cite{Sengupta:1994aa, Haggard:1986aa} such as Republic of Korea, Thailand and China.  These results commonly point to that the export commodity portfolio underlies different time-evolutions of the economic power of individual countries. 

\section*{Discussion}
\label{sec:discussion}

We have  investigated development patterns in the export commodity portfolios of countries by analyzing their correlation with the GDP.  Focussing first on the individual commodity categories, we have shown  that the share of  {\it Machinery and transport equipment} category exhibits positive cross-country and temporal correlations with  the GDP and  that two categories - {\it Food and live animals} and  {\it Crude materials} show  negative correlations.  These three categories, among 10 in the SITC, can be considered as relevant to economic power and development of countries at least in the studied period 1962 to 2000; Countries with larger share of the {\it Machinery} category are likely to have higher GDP than those with smaller share for  given year, and countries increasing its share  are likely to increase its GDP with time.  

Then we have shown that the whole export portfolio of countries are related to  to their economic growth via these GDP-correlated categories in relating.  With several clusters displaying distinct representative portfolios, we have shown that the transition to a cluster with its  representative export portfolios dominated by the {\it Machinery} category is necessary for the GDP growth. The increase of the global share of the {\it Machinery} category and the decrease of the share of the {\it Food} and {\it Crude materials} with time are identified in the world trade, which is the global environment specific for the studied period 1962 to 2000. Our results can be interpreted as that following such global trend in the export portfolio is necessary for the economic growth of individual countries, exemplified by Asian newly industrialized countries, while there are exceptions such as oil-producing countries.

Our study reveals a typical pattern in the change of the export portfolios  with the development level of countries. Deviations can be found for individual countries, the quantitative aspect of which is worthy of study. Not only the GDP, but also multiple factors including globalization and regional integration~\cite{RePEc:cpr:ceprdp:12387,BA87480587} as well as the population, tariff level, distance to other countries, and factor endowment~\cite{10.2307/2662292} may altogether influence the commodity composition of domestic production and international trade, which is not analyzed in the present study. The empirical analysis and mathematical modeling of such combinatorial influences can greatly deepen our understanding of the structure of international trade. Abrupt changes in the global environment such as economic crises, not considered in this work, should affect significantly the structure of international trade and its study is highly desirable. 

\section*{Methods}
\subsection*{Compiling data-sets}
\label{sec:dataset}

We use the export value data $F_p(c,t)$ in nominal thousands of US dollars with the  commodity classification following the SITC Rev. 2 available in the NBER-UN data-set~\cite{NBERw11040} and the GDP data $W(c,t)$ in constant US dollars (base 1996) from the data compiled by Gleditsch~\cite{GLEDITSCH01102002} based on the Penn World Table 6.1~\cite{pwt61}. Actually the units of $F_p(c,t)$ and $W(c,t)$ are irrelevant to our study for the normalized quantities in Eqs.~(\ref{eq:phipt}), (\ref{eq:phipct}), and (\ref{eq:gct}). We compare the full name of each country between the two data-sets to identify each country.

In the NBER-UN data-set, Russia and USSR appear together in year 1989, 1990, and 1991, and we take the sum of their export values as the export value of Russia in that period. Similarly, Gemany and German Federal Republic (GFR) appear together in 1989 and 1990, the export values of which are summed and taken as the export of GFR in that period. 

\section*{Data availability}
The datasets generated during  the current study are available in the GitHub repository~\cite{choi_github}.




\section*{Acknowledgements}
This work was supported by the National Research Foundation of Korea (NRF) grants funded by the Korean Government (No. 2019R1A2C1003486) and Inha University Research Grant (No. 59212).

\section*{Author Contributions}

S.-G.C. and D.-S.L. designed and performed  the research and wrote the  manuscript. 

\section*{Additional Information}
\textbf{Supplementary information} accompanies this paper. \\
\textbf{Competing Interests:} The authors declare no competing interests.


\makeatletter 
\setcounter{figure}{0}
\renewcommand{\thefigure}{S\@arabic\c@figure}
\setcounter{table}{0}
\renewcommand{\thetable}{S\@arabic\c@table}
\makeatother

\section*{\center \LARGE Supplementary Information}

\section*{GDP-correlated commodities at two-digit level}
\label{sec:twodigits}

\begin{figure*}
\includegraphics[width=\columnwidth]{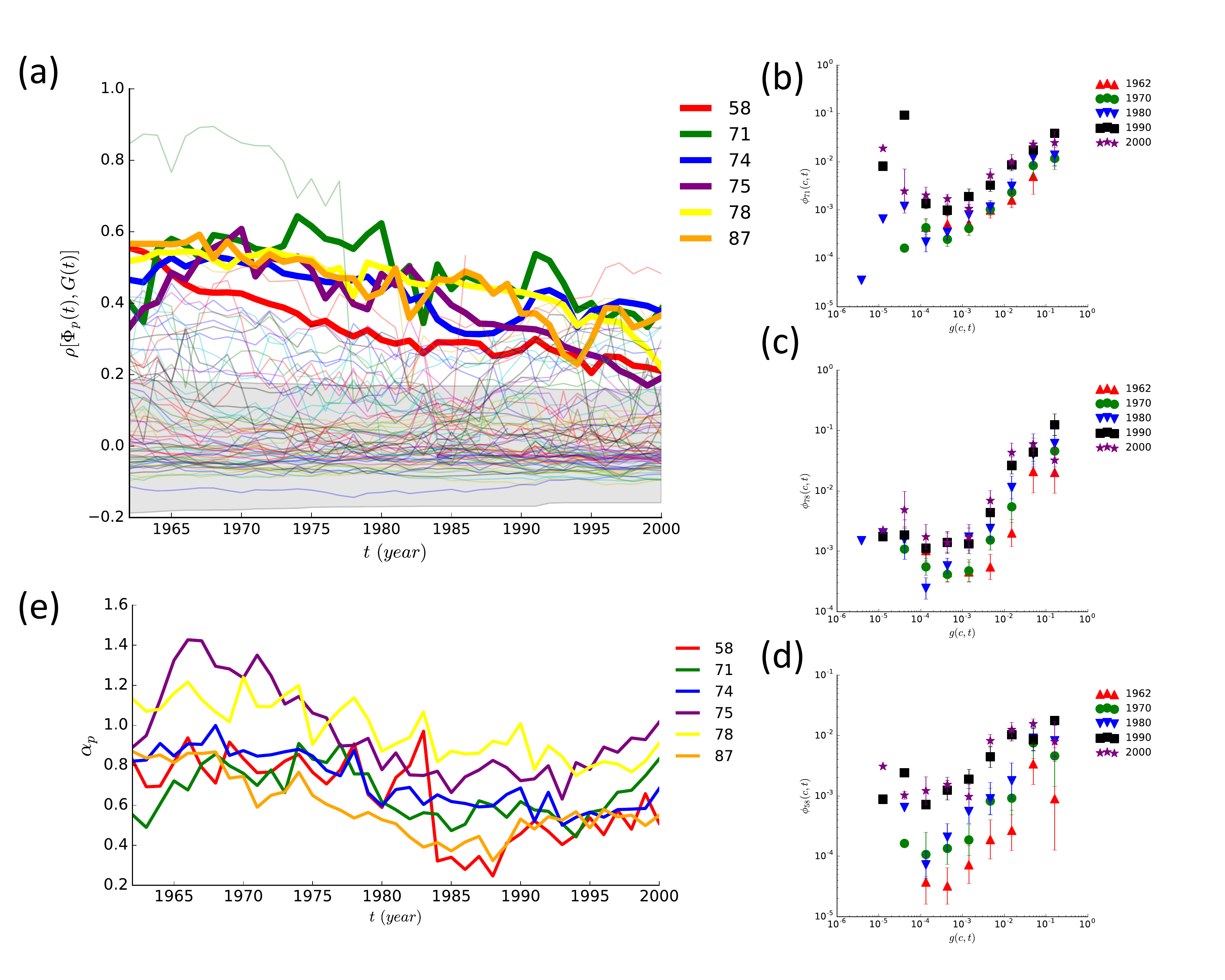}
\caption{
Correlation between the GDP profile and the export share profile with 93 commodity categories based on the first  two digits used. 
(a) The Pearson correlation coefficient in Eq.~(4) between the local export share profile of a commodity and the  GDP  profile as a function of time $t$. The range of $\rho$ with P $>0.05$ is shaded. Thick lines are used for the six categories showing correlations with $P<0.05$. 
 (b) Plot of the share $\phi_{71}(c,t)$ of the {\it Power-generating machinery and equipment} category versus the normalized GDP $g(c,t)$  for selected years. 
 (c) Plot of the share $\phi_{78}(c,t)$ of the {\it Road vehicles}  category versus the normalized GDP. 
 (d) Plot of the share $\phi_{58}(c,t)$ of the {\it Artificial resins and plastic materials, and cellulose esters etc}  category versus the normalized GDP. 
 (e) GDP-elasticity $\alpha_p(t)$ of the six GDP-correlated categories.  
}
 \label{fig:twodigits}
\end{figure*}

With  93 commodity categories  obtained by keeping  the first two digits of the SITC, we have investigated the relation between the shares of such refined commodity categories in the export values and the normalized GDP of countries and present the results in Fig.~\ref{fig:twodigits}.


\subsection*{Properties of the country clusters}
\label{sec:cluster_properties}


\begin{table*}
	\begin{tabular}{c|c||*{5}{p{0.7cm}}|*{7}{p{0.7cm}}}
		\hline\hline
		\multicolumn{2}{c||}{year ($t$)}& \multicolumn{5}{c|}{$1962$}&\multicolumn{7}{c}{$2000$}\\ 
		\hline
		\multicolumn{2}{c||}{Cluster($C$)} &$\mathit{0}$ & $\mathit{20}$ & $\mathit{2}$ & $\mathit{3}$ & $\mathit{76}$ & $\mathit{026}$ &$\mathit{20}$ & $\mathit{3}$ & $\mathit{36}$ & $\mathit{76}$ & $\mathit{87}$ & $\mathit{8}$ \\ 
		\hline
		\multicolumn{2}{c||}{No. countries} & 31 & 28 & 17 & 9 & 25 & 18 & 10 & 17 & 6 & 31 & 15 & 7 \\ 
	\hline
		\multicolumn{2}{c||}{GDP($g(C,t)$)}& 0.11 & 0.18 & 0.013 & 0.026 & 0.67 & 0.036 & 0.0029 & 0.034 & 0.041 & 0.61 & 0.23 & 0.0054 \\
\hline
		\multirow{10}{*}{\shortstack[1]{Export\\ commodity\\portfolio}} 
		& $\phi_0(C,t)$ & 0.618 &0.320 &0.076 &0.018  &0.108 &0.282 &0.132 &0.017 &0.069 &0.049 &0.056 &0.122 \\
		&$\phi_1(C,t)$ &0.018 &0.025 &0.001 &   0.000 &0.017 &0.016 &0.019 &0.001 &0.002 &0.009 &0.005 &0.022 \\
		&$\phi_2(C,t)$ &0.164 &0.414 &0.698 &   0.032 &0.105 &0.186 &0.486 &0.011 &0.060 &0.030 &0.036 &0.015 \\
		&$\phi_3(C,t)$ &0.032 &0.097 &0.034 &   0.932 &0.042 &0.149 &0.048 &0.873 &0.458 &0.039 &0.053 &0.002 \\
		&$\phi_4(C,t)$ &0.023 &0.020 &0.017 &   0.000 &0.008 &0.011 &0.009 &0.000 &0.001 &0.003 &0.004 &0.001 \\
		&$\phi_5(C,t)$ &0.024 &0.016 &0.006 &   0.001 &0.077 &0.048 &0.023 &0.033 &0.062 &0.101 &0.042 &0.005 \\
		&$\phi_6(C,t)$ &0.052 &0.080 &0.142 &   0.010 &0.252 &0.164 &0.079 &0.033 &0.219 &0.138 &0.189 &0.084 \\
		&$\phi_7(C,t)$ &0.044 &0.011 &0.017 &   0.003 &0.305 &0.073 &0.067 &0.020 &0.037 &0.496 &0.260 &0.051 \\
		&$\phi_8(C,t)$ &0.020 &0.008 &0.006 &   0.001 &0.071 &0.032 &0.127 &0.008 &0.025 &0.107 &0.340 &0.684 \\
		&$\phi_9(C,t)$ &0.004 &0.010 &0.004 &   0.002 &0.015 &0.038 &0.011 &0.003 &0.068 &0.029 &0.015 &0.014 \\
		\hline\hline
	\end{tabular}
	\caption{The number of countries, the normalized GDP, and the export commodity portfolio of five clusters in 1962 and seven clusters in 2000, as identified in Fig. 4 (a), are shown. 	}
	\label{table:clusterportfolio}
\end{table*}


\begin{table*}
	\begin{tabular}{c||ccccccc}
	   \diagbox{$C^{(1962)}$}{$C^{(2000)}$}& $\mathit{026}$  & $\mathit{20}$ & $\mathit{3}$ & $\mathit{36}$ & $\mathit{76}$ & $\mathit{87}$ & $\mathit{8}$ \\
	     		\hline\hline	
		$\mathit{0}$ & 0.377 & 0.594 & 1.046 & 0.727 & 0.754 & 0.709 & 0.844\\
		$\mathit{20}$& 0.263 & 0.246 & 0.927 & 0.588 & 0.694 & 0.632 & 0.816 \\
                 $\mathit{2}$ & 0.570 & 0.265 & 1.092 &0.775 & 0.834 & 0.783 & 0.967 \\
                 $\mathit{3}$ & 0.861 & 1.013 & 0.076 & 0.531 & 1.039 &0.995 & 1.162 \\
                 $\mathit{76}$ & 0.336 & 0.488 & 0.918 & 0.505 & 0.247 & 0.296 & 0.695 \\	
	\end{tabular}
	\caption{The Euclidean distance between the average portfolios of the clusters of 1962 and 2000.}
		\label{table:distance}
\end{table*}

Some properties of each cluster, including the average share of each category are given in  Table~\ref{table:clusterportfolio}. The Euclidean distance between the average portfolios of each pair of clusters of 1962 and 2000 is shown in Table~\ref{table:distance}.

\subsection*{Time-evolution of the GDP of countries classified by their transition behaviors }

\begin{figure*}
\includegraphics[width=17cm]{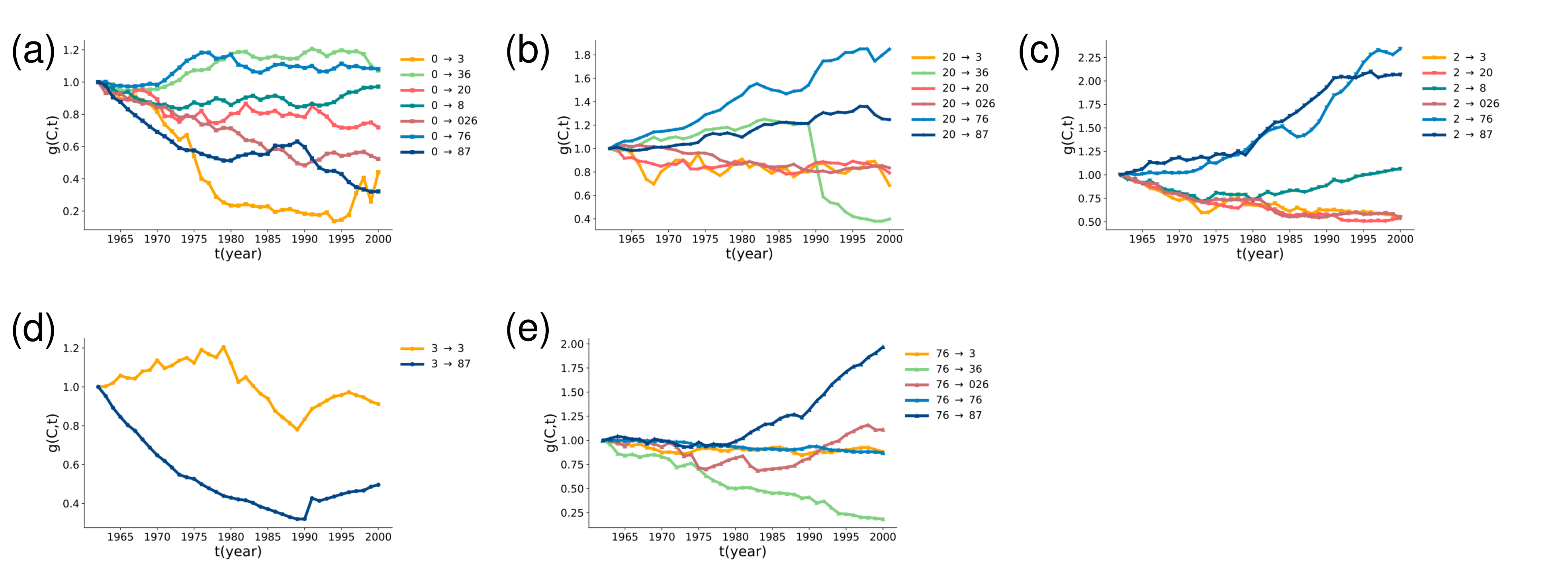}
\caption{
The time evolution of the normalized GDP of countries classified according to their transition between clusters in 1962 and 2000. For the countries transiting from a cluster $C$ in 1962 to a cluster $C'$ in 2000, we plot their  normalized GDP divided by that in 1962, $g(C\to C', t)/g(C \to C', 1962)$ as a function of time. The shape of data points varies with the cluster of 1962 and the color varies with the cluster of 2000. 
(a) The time evolution of the rescaled GDP $g(\mathit{0} \to C', t)/g(\mathit{0} \to C', 1962)$ for the countries which are in the cluster $\mathit{0}$ in 1962 and transit to a cluster $C'$ in 2000. 
(b) Plots of $g(\mathit{20} \to C', t)/g(\mathit{20} \to C', 1962)$ versus time $t$  for different clusters $C'$ in 2000. 
(c) $g(\mathit{2} \to C', t)/g(\mathit{2} \to C', 1962)$ versus time $t$.
 (d) $g(\mathit{3} \to C', t)/g(\mathit{3} \to C', 1962)$ versus time $t$.
 (e) $g(\mathit{76} \to C', t)/g(\mathit{76} \to C', 1962)$ versus time $t$.
}
\label{fig:individualcluster}
\end{figure*}

The time-evolution of the normalized GDP of all countries classified according to their transition between the clusters of 1962 and 2000 is shown in Fig.~\ref{fig:individualcluster}, which is an extended version of Fig. 5 and  consists of five panels each for the countries starting at the same cluster in 1962 and arriving at different clusters in 2000.


\end{document}